\documentclass{llncs}

\usepackage[utf8]{inputenc}
\usepackage{url}
\usepackage[british]{babel}
\usepackage{calc}

\usepackage{listings}
\title{%
  Automatic Coding Rule Conformance Checking Using Logic Programs
  \thanks{\footnotesize Work partially supported by Spanish Ministry
    of Industry PROFIT grants FIT-340005-2007-7 and
    FIT-350400-2006-44, Comunidad Autónoma de Madrid grant
    S-0505/TIC/0407 (PROMESAS), Ministry of Education and Science
    grant TIN2005-09207-C03-01 (MERIT/COMVERS) and EU IST FET grant
    IST-15905 (MOBIUS).}  \vspace{-1em} }

\author{
  Guillem Marpons\inst{1} \and
  Julio Mariño\inst{1} \and
  Ángel Herranz\inst{1} \and
  Lars-{\AA}ke Fredlund\inst{1} \and
  Manuel Carro\inst{1} \and
  Juan José Moreno-Navarro\inst{1,2}
}
\institute{
  \mbox{Universidad Politécnica de Madrid \hspace{3em} \and IMDEA Software}
  \email{\{gmarpons,jmarino,aherranz,lfredlund,mcarro,jjmoreno\}@fi.upm.es}
  \vspace{-1em}
}
\date{\today}

\begin{document}

\enlargethispage{1em} 

\maketitle

\begin{small}

%\subsubsection*{Abstract}\aftersect

  % Coding rules are customarily used to constrain the use (or abuse) of
  % certain constructions in a programming language.

  % Standard coding rule sets exist that target different languages and
  % application domains.
\bigskip\noindent
Some approaches to increasing program reliability 
% and maintainability
involve a disciplined use of programming languages so as to minimise
the hazards introduced by error-prone features.
This is realised by writing code that is constrained to a subset of
the \emph{a priori} admissible programs, and that, moreover, may use only
a % well-defined
subset of the language.  These subsets are determined by a collection of
so-called \emph{coding rules}.
Standard coding rule sets exist that target different languages
(e.g. MISRA-C for the C language or HICPP for C++) and application
domains.  Some organisations do set up their own coding rule sets.
%
% Additionally, organisations can set up their own internal rules.

  % However, these rules are usually written using a natural language,
  % which is intrinsically ambiguous, and which may hinder their automatic
  % application.

  % This short paper presents some early work in the context of an
  % European project aiming at defining a framework to formalise and check
  % for coding rule conformance using logic programming.

A major drawback of actual coding rule sets is that they are written
in natural language, which bears ambiguity and undermines 
% current industry-directed efforts 
any effort
to enforce them automatically.
This work aims at defining a framework to formalise coding rules and
check for conformity with them, using logic programming.
%
% It is being undertaken as 
It is 
part of the Global GCC project
(\texttt{http://www.ggcc.info/}),
% a consortium of EU industrial
% corporations and research labs funded under the ITEA Programme and
an ITEA funded EU programme intended to enrich 
% the capabilities of 
the GNU Compiler Collection with advanced project-wide 
% compile-time
analysis capacities.

  % We show how a certain class of rules -- \emph{structural} rules -- can
  % be reformulated as logic programs.

The overwhelming diversity of rules (they range from being trivially
enforceable
%
% (MISRA-C 20.4: ``\emph{do not use the \texttt{malloc()} function}'')
%
to expressing non-com\-pu\-ta\-ble properties)
%
% (HICPP 3.1.9: ``\emph{behavior must be implemented by only one member
%   function}'').
%
% Without renouncing to work on other rule types, 
has obliged us to focus first
on a particular class that we have termed \emph{structural rules}:
those which deal with static entities in the code (classes, member
functions, etc.) and their properties and relationships (inheritance,
overriding, etc.)
We have identified a significant number of rules of this kind that can
be statically checked, being at the same time more interesting than
those purely syntactic.

  % This provides both a framework for formal specification and also for
  % automatic conformance checking using a Prolog engine.

  % Some real examples, including Prolog code formalising them, are shown
  % and discussed.

Rules are formalised using first order logic:
% , as it is expressive
% enough to easily capture the meaning of short sentences in
% well-defined domains.
relationships between program entities are encoded as facts 
(thus giving an abstract description of the program) 
and a formula is generated for every coding rule.
When these, together, are inconsistent, the program violates the
coding rule.
We automate this process by generating a program-dependent set of
Prolog facts and program-independent predicates
which describe rule violations.
For example, a violation of rule 3.3.15 of HICPP, which reads
``\emph{ensure base classes common to more than one derived class are
  virtual}'', is codified as:

\begin{minipage}{\linewidth}
\begin{lstlisting}[basicstyle=\ttfamily\scriptsize\selectfont]
   violate_hicpp_3_3_15(A,B,C,D) :- class(A), class(B), class(C), class(D), 
        B \= C, direct_base_of(A,B), direct_base_of(A,C), 
        base_of(B,D), base_of(C,D), \+ virtual_base_of(A,C).
\end{lstlisting}
\end{minipage}

% \noindent
% \begin{tabular}{p{0.45\linewidth}@{~~}p{0.48\linewidth}}
%   \hspace{1.1em}
%   We automate this process by generating a Prolog
%   program containing such set of facts, and clauses defining a
%   predicate which describes a rule violation.
% %
%   The predicate at the right codifies the violation of rule 3.3.15 of
%   HICPP, which reads ``\emph{ensure base classes common to more than
%     one derived class are virtual}''.
% &
% \vspace*{-1\baselineskip}
% \begin{lstlisting}[basicstyle=\ttfamily\footnotesize\selectfont]
% violate_hicpp_3_3_15(A,B,C,D) :-
%     class(A), class(B), 
%     class(C), class(D), B \= C,
%     direct_base_of(A,B), 
%     direct_base_of(A,C), 
%     base_of(B,D), base_of(C,D), 
%     \+ virtual_base_of(A,C).
% \end{lstlisting}
% \end{tabular}

Successful queries to this predicate pinpoint infringements of the
rule and the answer substitutions identify a source of the violation.
% A closely related approach, applied to a different realm, has been
% followed by~\cite{fabry:2004:lid,blewitt:2001:avj-short}.
%
  % The code obtained is used to speculate on the possible challenges on
  % efficiency and the role that certain extensions to logic programming
  % can play.

As rule-writers may not be proficient in  Prolog, 
%rules are not formalised directly in Prolog: 
we provide a user-friendly domain-specific
language (DSL) that also increases expressiveness by, e.g., allowing
quantification over some specific domains or providing facilities for
defining closures.
At the DSL core there is a set of predefined predicates describing
(structural) program properties, such as those used in the above rule,
that are gathered during the compilation process.

% Negation
% is currently handled by suitably adding calls to enumerating
% predicates which guarantee the instantiation state required by
% Prolog's negation as failure.  This is a point upon which we want to
% improve by using constructive negation.

%% Since Prolog syntax may be alien to many rule-writers
%% % (and far away from the source code to be checked) 
%% and Prolog capabilities may exceed what is needed in most cases,
% (while its expressiveness falls short in others),
%
% Nonetheless, pure Prolog is not appropriate for the user writing its own
% rules. She can not be used to its syntax and many of its features are
% irrelevant to our setting.  For this reason 
%
% We plan to provide a user-friendly domain-specific
% language~\cite[Section 4]{marpons:2007:acr} which works around Prolog peculiarities.
% %% for more details, including hints to deal with negation, quantification on
% %% specific domains and closures, and its translation to Prolog.) 
% A set of predefined predicates describing (structural) program
% properties, gathered by a compiler, sits at the core of this DSL.
% %which are used to codify rules. 
% %Some of them are shown in the next example.

\end{small}

\end{document}